# Amorphous carbon film deposition on inner surface of tubes using atmospheric pressure pulsed filamentary plasma source

**Ramasamy Pothiraja, Nikita Bibinov and Peter Awakowicz**
Institute for Electrical Engineering and Plasma Technology, Ruhr-Universität Bochum, 44801 Bochum, Germany
E-mail: ramasamy.pothiraja@rub.de, nikita.bibinov@rub.de and awakowicz@aept.rub.de

**Abstract.** Uniform amorphous carbon film is deposited on the inner surface of quartz tube having the inner diameter of 6 mm and the outer diameter of 8 mm. A pulsed filamentary plasma source is used for the deposition. Long plasma filaments (~ 140 mm) as a positive filamentary discharge are generated inside the tube in argon with methane admixture. FTIR-ATR, XRD, SEM, LSM and XPS analyses give the conclusion that deposited film is amorphous composed of non-hydrogenated $sp^2$ carbon and hydrogenated $sp^3$ carbon. Plasma is characterized using optical emission spectroscopy, voltage-current measurement, microphotography and numerical simulation. On the basis of observed plasma parameters, the kinetics of the film deposition process is discussed.

## 1. Introduction

Uniform thin film deposition on the inner surface of curved three-dimensional objects is one of the challenging processes in modern plasma technology. Several research groups have reported a systematic development of such film deposition processes like coating of inner surface of tubes and bottles [1-11]. Mostly microwave or RF driven ICP, CCP and jet based microplasma or magnetron plasma sources were used for their studies. Generally, in several methods used for coating films on inner surface of tubes, precursors are decomposed in the confined electrode region where the plasma is active, the chemically active species are transported inside the tube and film coating is carried out by gas flow. In this case, because of polymerization and recombination reactions during the transport process, the nature of the chemically active species (constituent of polymer film) is different at different places along the axis of the tube. This phenomenon could reduce film uniformity along the axis of the tube. In our method of film coating on inner surface of tubes, a long filament of plasma is generated inside the tube in a gas containing a precursor as an admixture [12]. The plasma filament, which is thinner than the diameter of the tube to be coated, is active for long distance in the region of film coating. This plasma filament can ionize and/or dissociate precursor molecules. Using this method, chemically active species can be generated everywhere along the axis of the tube within close vicinity of inner surface of the tube. In this way, the differences in the nature of depositing chemically active species at different places along the axis of the tube can be reduced and films with better uniformity can be deposited. In addition to this, film deposition is supported by ion bombardments, which results in the formation of high quality film.

In the previous publication [12], we have reported our plasma source characterization and the model used for plasma characterization. In the present studies, the plasma source has been modified in order to increase the distance between the electrodes as well as to reduce the region which is covered by the grounded electrode. Characterization of deposited film, determination of plasma parameters, and simulation of chemical kinetics are carried out for this modified plasma source. In this paper, we report the following, 1. configuration of our modified plasma source, 2. film deposition with two different positions of driven electrode, 3. characterization of the film deposited on the inner surface of the tube, 4. plasma parameters, 5. simulation of chemical kinetics on the basis of determined plasma parameters, 6. correlation of film properties with the plasma parameters and chemical kinetics, and 7. mechanism of film deposition processes.

## 2. Experimental setup and simulations

### *2.1. Experimental setup*

Experimental setup is similar to one reported previously [12], except dimension and location of the grounded electrode (figure 1). A 10 mm long copper tube is used as a grounded electrode. For all the experiments reported in this article, the grounded electrode is fixed at 140 mm apart from the spike of the driven electrode. We performed two sets of experiments. First set of experiments are carried out in argon and methane mixture (Ar, 99.87%; $CH_4$, 13%, flow rate, 2.4 slm) to deposit carbon based film on inner surface of the tube. It is carried out at two different positions of driven electrode. Films deposited at this condition are characterized using various surface analysis techniques. Second set of experiments are carried out with argon, methane and nitrogen gas mixture (Ar, 99.85%; $CH_4$, 11%, $N_2$, 0.04%; flow rate, 2.4 slm) for plasma characterization. Emissions of nitrogen molecules and nitrogen ions are used for the determination of plasma parameters. Plasma parameters determined at this condition are considered to be the same as the plasma parameters during film deposition process in first set of experiments. The effect of absence of nitrogen on plasma parameters is balanced by increasing the methane quantity. This fact (balancing of plasma parameters) is confirmed from the similar simulated EVDFs for both sets of experiments (with/without nitrogen). It is also confirmed by measuring argon emission spectra at various places along the axis of the tube in both cases. Argon emission intensities are very close to each other in both cases and also have the similar trend along the axis of the tube. Analytical instruments used for plasma characterization are same as used for previous characterization [12], except that the spatial resolution of CCD camera is about 10 μm for the objective used for the plasma volume measurement in this study.

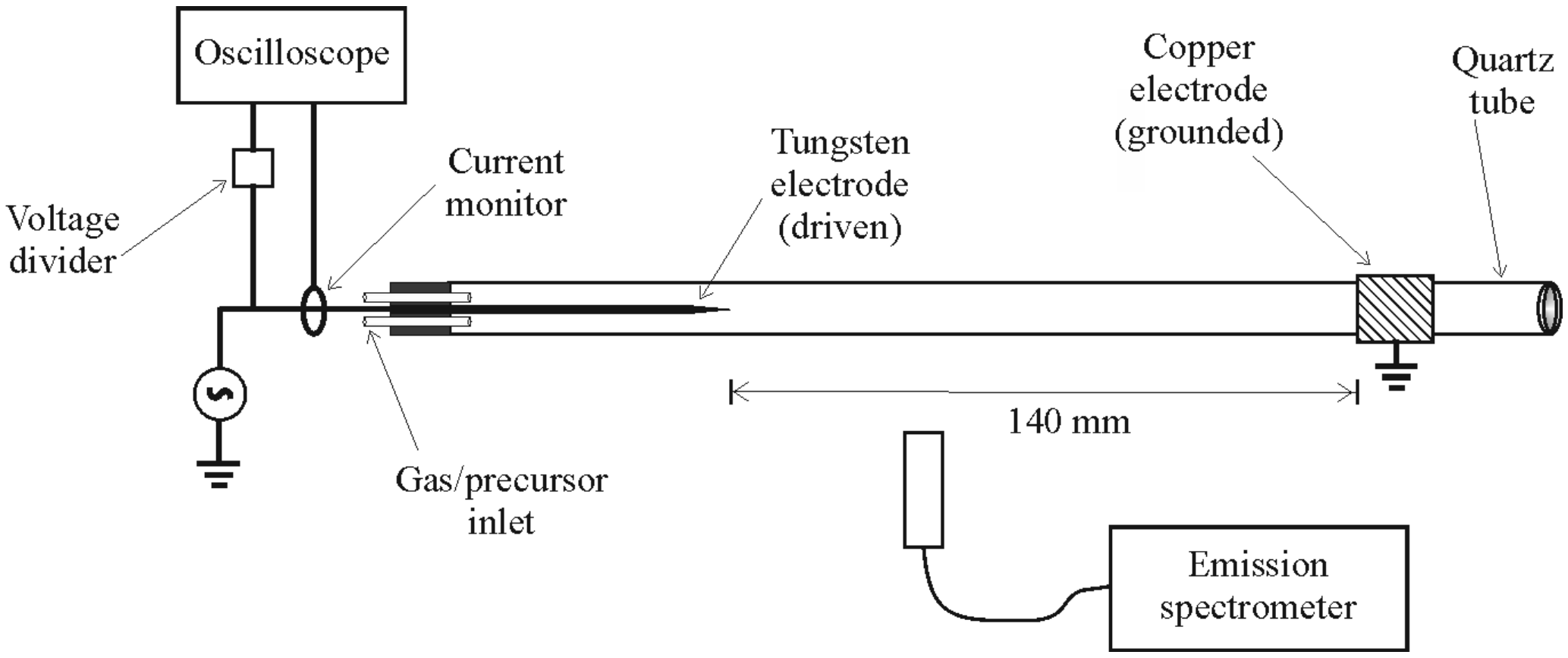

**Figure 1.** Schematic view of the experimental setup.

Since the film is coated on the inner surface of the tube, we do not have the possibility of recording FTIR-ATR, XPS and XRD spectra for this film. Hence, the film is coated also on quartz plates as follows. Quartz plates of dimensions 0.1-0.23 mm × 2-3 mm × 25 mm and /or 1 mm × 2-3 mm × 25 mm are placed inside the quartz tube at different places in the region between the powered and the grounded electrodes. Film deposition is carried out on these quartz plates. The film coated plates are used for FTIR-ATR, XRD and XPS spectral measurements. A FTIR spectrometer (BRUKER VECTOR 33) with a resolution of 0.3-1 $cm^{-1}$ is used. Colour 3D laser scanning microscope (LSM) (Keyence, pitch resolution, 100 nm; Z measurement distance, 13 μm; XY calibration, 46 nm/pixel; Z calibration, 1 nm/digit) is used for surface profile measurement. LSM and scanning electron microscope (SEM) (LEO Gemini SEM 1530 electron microscope) are used to obtain the cross sectional images of film. Film thickness is determined from the cross-sectional image of the film in the tube [13]. These values obtained from the images of LSM and SEM are in accordance with each other. A few atomic layer of gold film is coated on the samples used for SEM measurements for conductivity. Information about amorphous nature of the film is obtained using X-ray diffraction analysis (Bruker D8 Advanced AXS diffractometer; Cu-Kα radiation (1.5481 Å); position sensitive detector (PSD); films are analyzed in the $\theta$-$2\theta$ geometry and measured in Bragg-Brentano geometry at room temperature; constant divergence slit is 5 mm; Ni-filter is used; step 0.02°; 90 seconds per point). Elements present in the film are determined using EDX spectra. EDX spectra are obtained from the spectrometer attached to the SEM instrument. Chemical nature of the elements is analyzed using XPS spectra (Physical Electronics, PHI 5000 *VersaProbe*, Al-Kα radiation). For a survey spectrum, it is scanned in the binding energy range of –2 to 1198 eV with the pass energy 187.85 eV; the energy step 0.5 eV; number of cycles 5. For the core level line measurement, the pass energy is kept as 23.5 eV with the energy step of 0.05 eV. Temperature of the tube is measured using digital thermometer attached with K type thermocouple (Greisinger electronic, model GTH 1150). For this purpose, plasma is ignited for 10 minutes, and then it is switched off. Immediately after switching off the plasma, temperature is measured by keeping the thermocouple in contact with the tube. The error in measurement due to the response time of thermocouple is corrected with the temperature decay profile. Absorption spectra are obtained using a deuterium lamp as a light source. Ocean optics emission spectrometer is used for transmitted light measurements.

*2.2. Determination of gas temperature, EVDF, excitation rate constants and electron density*

Optical emission spectroscopy (OES), voltage-current measurement, microphotography and numerical simulations are used for the determination of gas temperature in the filament and the plasma parameters. The details of applied diagnostic methods are described in our previous publication [12]. Therefore, it is described only shortly as follows. Gas temperature in active plasma volume is one of the most important parameters, because of its influence on gas density in plasma and on the rate constants of chemical reactions. It is determined from the rotational temperature of diatomic molecules namely, $N_2$ and CN, from the emission of $N_2$(C–B, 0–0) and CN(B–X, 0–0).

The relative intensity of $N_2$(C–B, 0–0) with respect to the intensity of $N_2^+$(B–X, 0–0) is used for the determination of EVDF. For this purpose, the relative intensity of $N_2$(C–B, 0–0) with respect to the intensity of $N_2^+$(B–X, 0–0) is simulated for various EVDFs for our experimental conditions by numerical solution of the Boltzmann equation in local approximation and varied electric field applying the program code "EEDF" developed by Napartovich *et al* [14]. Finally, by comparing the experimentally determined relative intensities ($I_{N_2^+(B-X)}/I_{N_2(C-B)}$) with the simulated relative intensities ($I_{N_2^+(B-X)}/I_{N_2(C-B)}$) for various EVDFs, the actual EVDF is determined.

Using the normalized EVDF and the known collisional cross section $\sigma_{exc}$ (cm$^2$) for electron impact excitation [15], we calculate the rate constants $k$ (cm$^3$·s$^{-1}$) for electron impact excitation of $N_2$, Ar and $CH_4$ using the equation (1):

$$k = 4\pi\sqrt{2}\int_0^\infty f_v(E) \cdot \sqrt{\frac{2C}{m_e}} \cdot E \cdot \sigma_{exc}(E)\, \mathrm{d}E, \qquad (1)$$

where, $m_e$ is the mass of electron (g), $E$ is the kinetic energy of electrons (eV) and $C = 1.602 \times 10^{-12}$ erg· eV$^{-1}$.

The electron density ($n_e$, cm$^{-3}$) is determined using the equation (2) from the measured absolute intensity of $N_2$(C-B, 0-0) emission ($I_{N_2(C-B)}$, phot·cm$^{-3}$·s$^{-1}$), the nitrogen density ([$N_2$], cm$^{-3}$), the electron impact excitation rate constant for $N_2$(C-B, 0-0) emission ( $k_{N_2(C)}$, cm$^3$·s$^{-1}$), the contribution of excitation of $N_2$(C) by collision with argon metastables ($K_{N_2(C)}^{Ar_{met}}$) [12], the contribution of the quenching of $N_2$(C) by argon ($Q_{N_2(C)}$), the plasma volume ($V_p$, cm$^3$), the value of fraction of time in which plasma is active ($t_f$), and the geometrical conditions during the measurement of emission spectrum ($g_f$).

$$n_e = \frac{I_{N_2(C-B)}}{[\mathrm{N_2}] \cdot (k_{N_2(C)} + K_{N_2(C)}^{Ar_{met}}) \cdot Q_{N_2(C)} \cdot V_p \cdot t_f \cdot g_f} \qquad (2)$$

*2.3. Temporal and spatial distribution of the gas temperature as well as fluxes of chemically active species*

In order to determine the temporal and spatial distribution of gas temperature as well as fluxes of chemically active species from the plasma filament towards the surface of the tube, we numerically solve the equations for thermal conductivity (3) and diffusion (4) for a cylindrical symmetry [16, 17]:

$$\frac{\partial T}{\partial t} = \nabla \cdot (\alpha(T)\nabla T) \qquad (3)$$

$$\frac{\partial [M]}{\partial t} = \nabla \cdot (D(T)\nabla [M]) \qquad (4)$$

where,

$$\alpha(T) = \frac{\lambda_{\mathrm{conduction}}}{\rho C_p}$$

$\alpha(T)$ - thermal diffusivity

$\lambda_{\mathrm{conduction}}$- thermal conductivity

$\rho$ - density

$C_p$ - specific heat capacity

$D$ - diffusion coefficient

## 3. Results and Discussion

The pulsed positive filamentary discharge is ignited [18] in the mixture of argon and methane. The long filament of plasma generated along the axis of the tube during this discharge has a diameter of about 200 µm. The duration of the positive filamentary discharge is around 160 ns. During plasma operation, the profile and position of filaments in the tube are changing with the frequency of several Hz. Because of this fact (profile and position of filaments are stationary for about 100 ms) and the pulse frequency is 22 kHz, about 2000 filaments in series have the same profile and position. Films deposited at this condition are characterized using surface analysis techniques. For the films deposited in both experimental conditions (which differ in the position of driven electrode), nature of the elements, chemical composition and surface morphology are the same. However, variation of film thickness along the axis of tube is different. The details of film analysis are discussed below.

### *3.1. Characterization of deposited film*

LSM images of the surface of the films deposited on quartz plates show some roughness (Rz = 8 nm) (figure 2, right). Cross sectional images of the film deposited on quartz tube are obtained using LSM and SEM (figure 2, left). From these images, film thickness is determined at 65 mm distant from the spike. By keeping this value as a reference, film thickness at various places along the axis of the tube is determined using optical thickness of the film on the basis of Beer-Lambert law. From the known values of film thickness and duration of film deposition, the film growth rate is determined. UV-Vis absorption spectra (figure 3, left) of the films in the entire region between the electrodes show similar profiles of absorption with $\lambda_{max}$ at around 220 nm indicating the uniform nature of the deposited materials, composed of $sp^2$ carbon with a short range of conjugation [19].

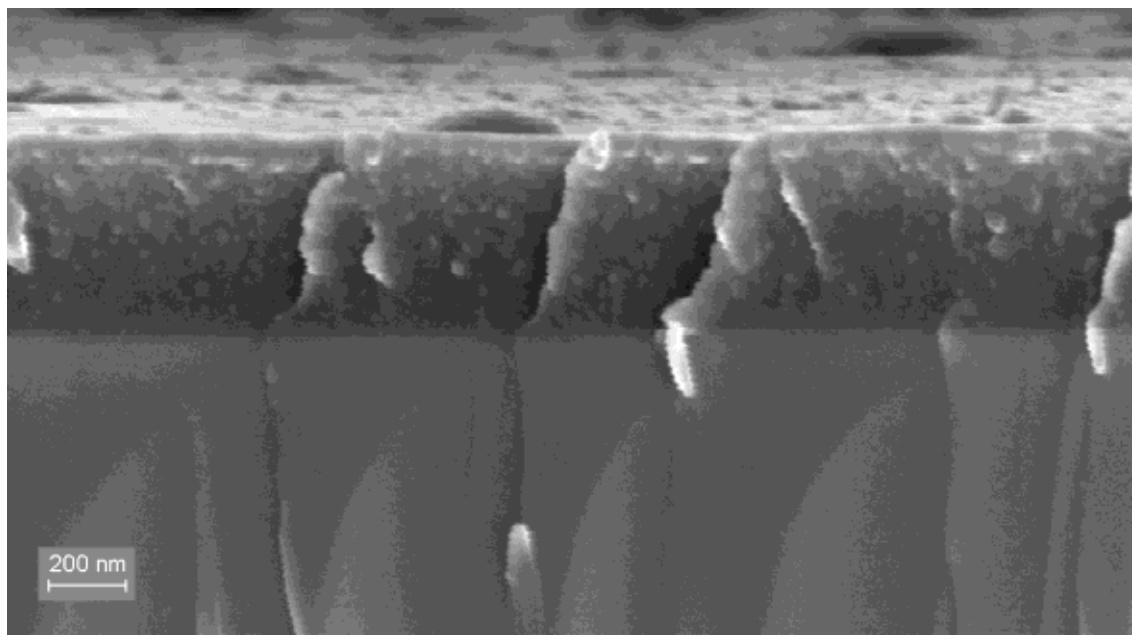

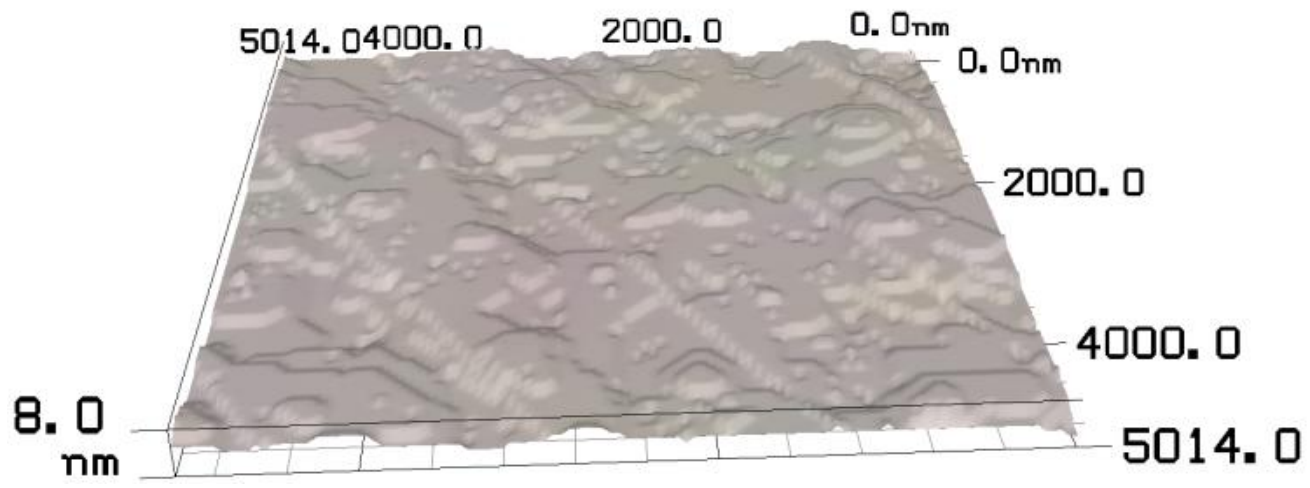

**Figure 2.** SEM image of the film cross-section (left) and LSM image of the film surface (right) at middle region in between electrodes.

For the determination of optical density, wavelength 350 nm is chosen in order to have both absorbance as well as transmittance for the films in all range of thicknesses. The following figure (figure 3, right) shows the optical density as well as growth rate of films deposited in two different experimental conditions. In the first experimental condition, the driven electrode is placed on the axis of the tube at the radial centre. In this case, the filaments ignited from the spike of driven electrode are stochastically distributed in the tube cross section and touch the tube surface only within the grounded electrode region. It results the maximum film growth rate at the middle region in between electrodes (figure 3, right, -○-). In the second experimental condition, the driven electrode with spike is shifted 0.3 mm from the radial centre of the tube towards the surface of the tube. In this case, all filaments approach the tube surface at the same point (at about 20 mm from the spike) and pass on tube surface for a distance of about 20 mm. From this point till the grounded region, the filaments are stochastically distributed around the axis of the tube. This causes increase in the deposition rate by a factor of about three, in the region, where the filament touches the surface of tube (figure 3, right, -■-). Film thickness is almost constant from about 40 to 120 mm from the spike. In both experimental conditions, the deposition rate is very low shortly before the grounded area. The deposition rate within the grounded electrode region is almost two orders of magnitude higher than the deposition rate in the area in between the electrodes. The increase in film growth rate near the spike, caused by the contact of plasma filament on the surface of tube, is important for the deposition of film on inner surface of tubes of different radii and will be investigated in future. Also, position of the filaments in the tube and therefore deposition rate can be controlled by placing a grounded conductor (e.g. wire) on or near the outer surface of tube.

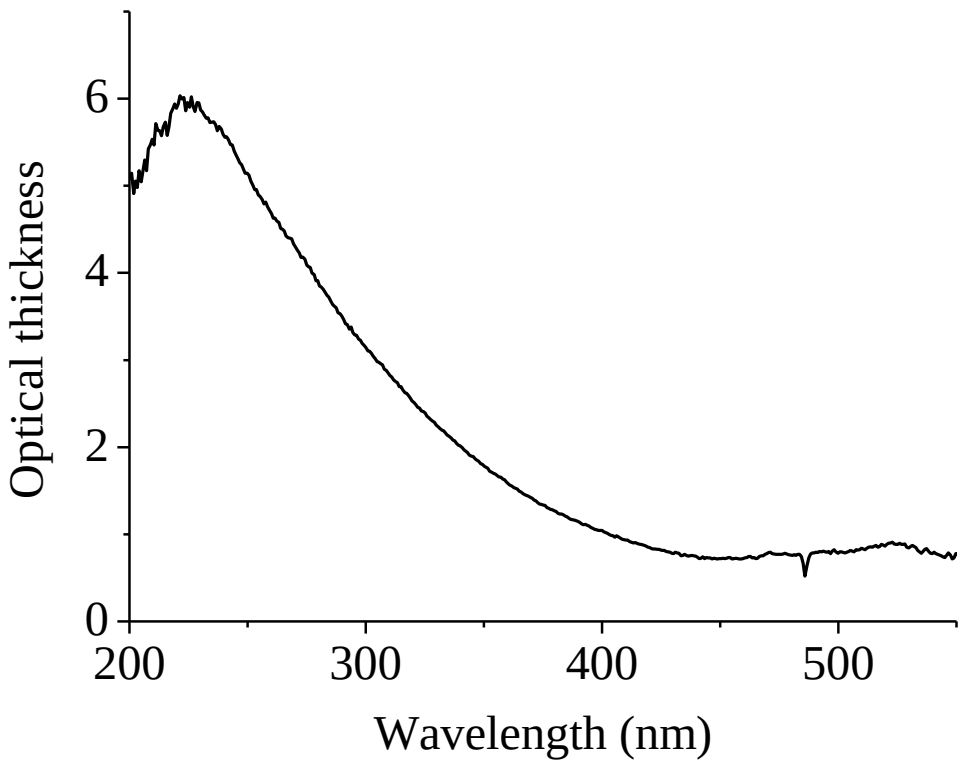

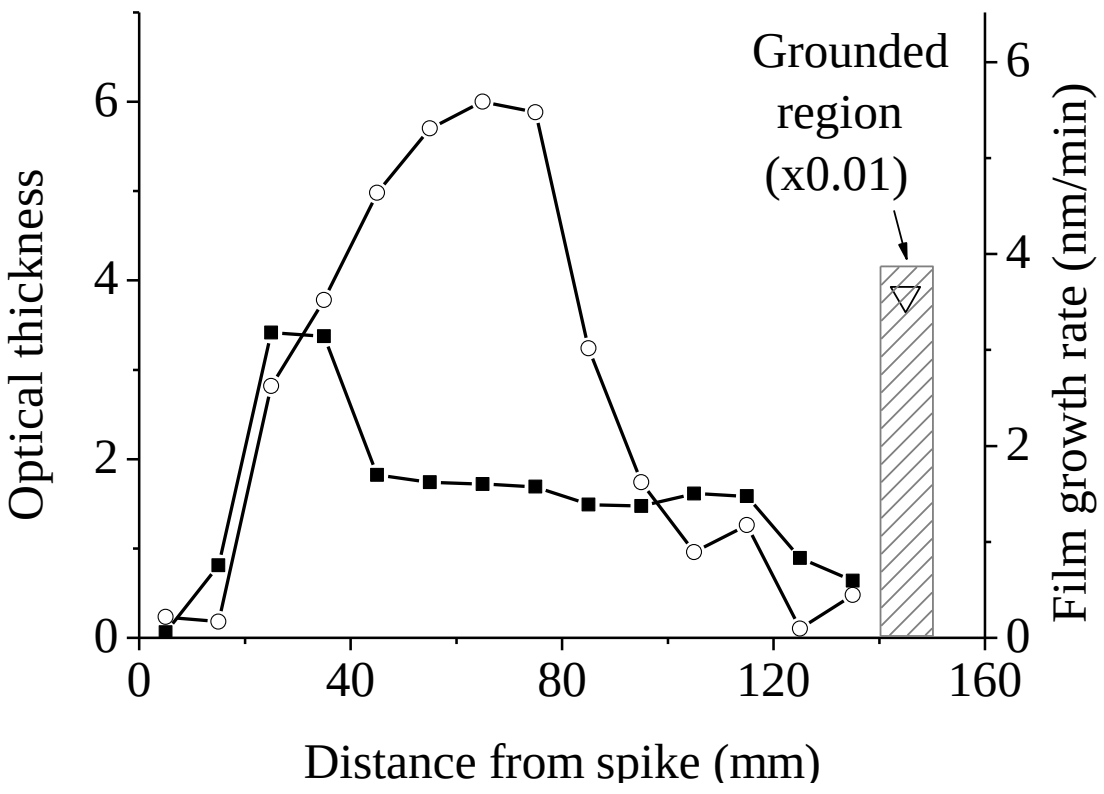

**Figure 3.** Left: UV-Vis absorption spectrum of the film; Right: Optical density of the films along the axis of the tube at 350 nm (-○-, electrode at the radial centre of the tube; -■-, electrode at 0.3 mm below the radial centre of the tube; ∇, value in the grounded electrode region is divided by 100).

FTIR spectra are measured in ATR mode with diamond crystal for the films deposited on quartz plates. Before measurement of the sample, background measurement in atmospheric air is carried out in the ATR mode. After this background measurement, FTIR spectra in the ATR mode are measured for the film coated quartz plate. Background corrected spectrum of film coated plate is considered as a spectrum of coated film [20, 21]. FTIR-ATR spectrum of the film on a quartz plate is shown in figure 4. There is a pseudo-high transmittance peak in the background corrected FTIR-ATR spectrum of film because of the presence of $CO_2$ in background air and its absence in the film. Hence, the peak at 2350 $cm^{-1}$ is not shown in the figure 4 for clarity. Analysis of FTIR-ATR spectra of the deposited films indicates the presence of aliphatic C-H groups in the film through their characteristic absorption. All the spectra have the peaks at 2954, 2926 and 2870 $cm^{-1}$ corresponding to $CH_3$ asymmetric stretch, $CH_2$ asymmetric stretch and $CH_2$ symmetric stretch, respectively [22-25], indicating the presence of hydrogenated $sp^3$ carbon in the film (figure 4). Presence of strong peak at around 1650 $cm^{-1}$ shows the presence of double bonded $sp^2$ carbon on the surface of the film [26, 27]. Absence of peak for stretching of ($sp^2$)C-H above 3000 $cm^{-1}$ indicates that $sp^2$ carbons are not hydrogenated. Hydrogen in the film is mainly attached to the $sp^3$ hybridized carbon. The relative intensity of "($sp^3$)C-H" absorption with respect to intensity of "($sp^2$)C=C(or O, N)" absorption varies slightly for the films on the quartz plates, which are placed at different places inside the tube between the electrodes during deposition. Sharp peaks at 1455 $cm^{-1}$ and 1375 $cm^{-1}$ are due to the bending vibrations of $CH_2$ and $CH_3$ groups respectively.

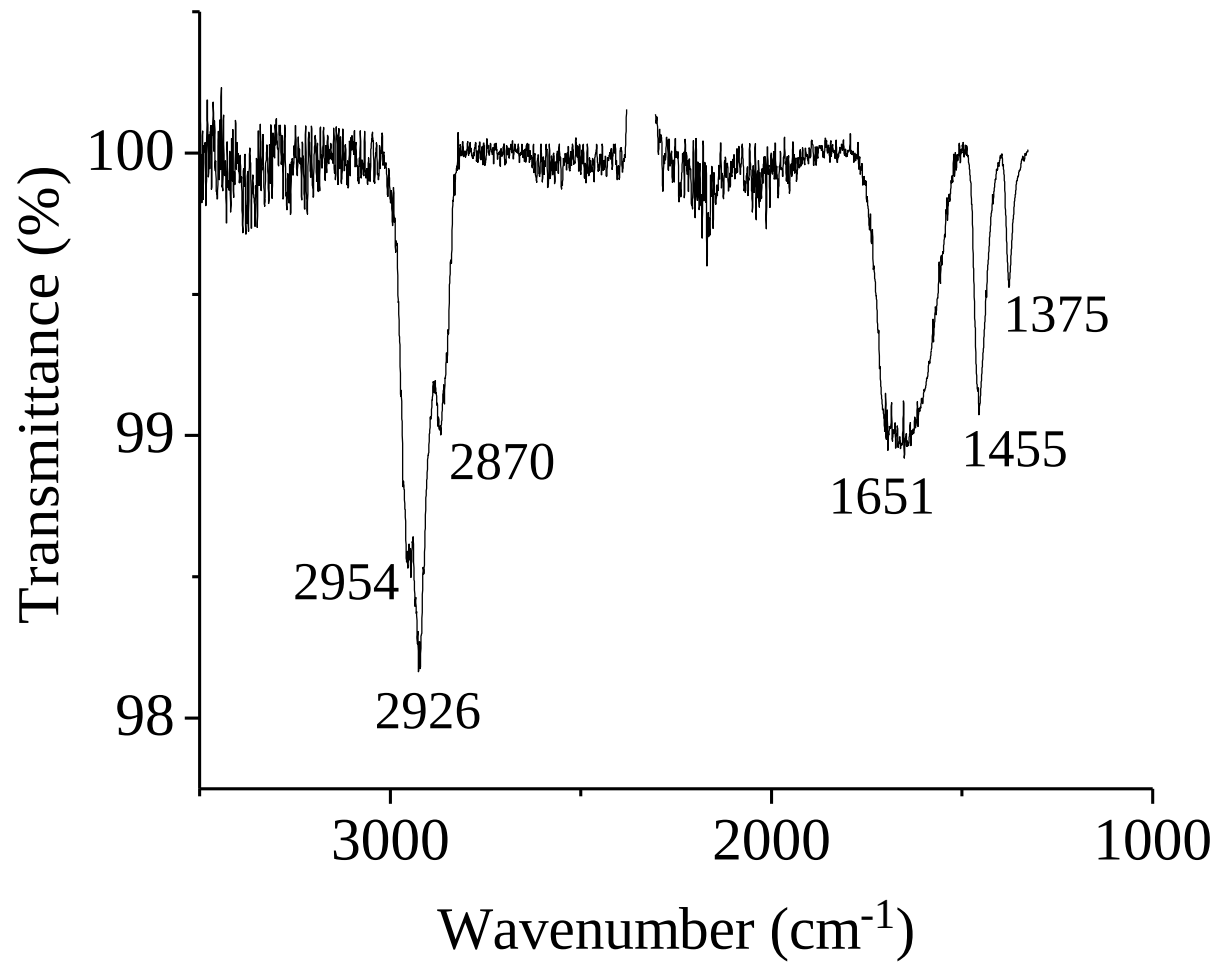

**Figure 4.** FTIR-ATR spectrum of the film deposited on the quartz plate placed in the middle region in between the electrodes during the pulsed positive filamentary discharge (Ar, 99.87%; $CH_4$, 0.13%; total gas flow rate, 2.4 slm). The region below 1300 $cm^{-1}$ is not shown because of strong overlapping of absorption of quartz plate. Pseudo-high transmittance at around 2350 $cm^{-1}$ due to the presence of $CO_2$ in air during the background measurement is removed for clarity.

X-ray diffraction analysis is carried out in order to find out the crystallinity of film. It does not show any characteristic diffraction for graphene or diamond form of carbon. It only shows diffraction pattern characteristic for quartz substrate as shown in figure 5. It clearly indicates that the film is highly amorphous [28].

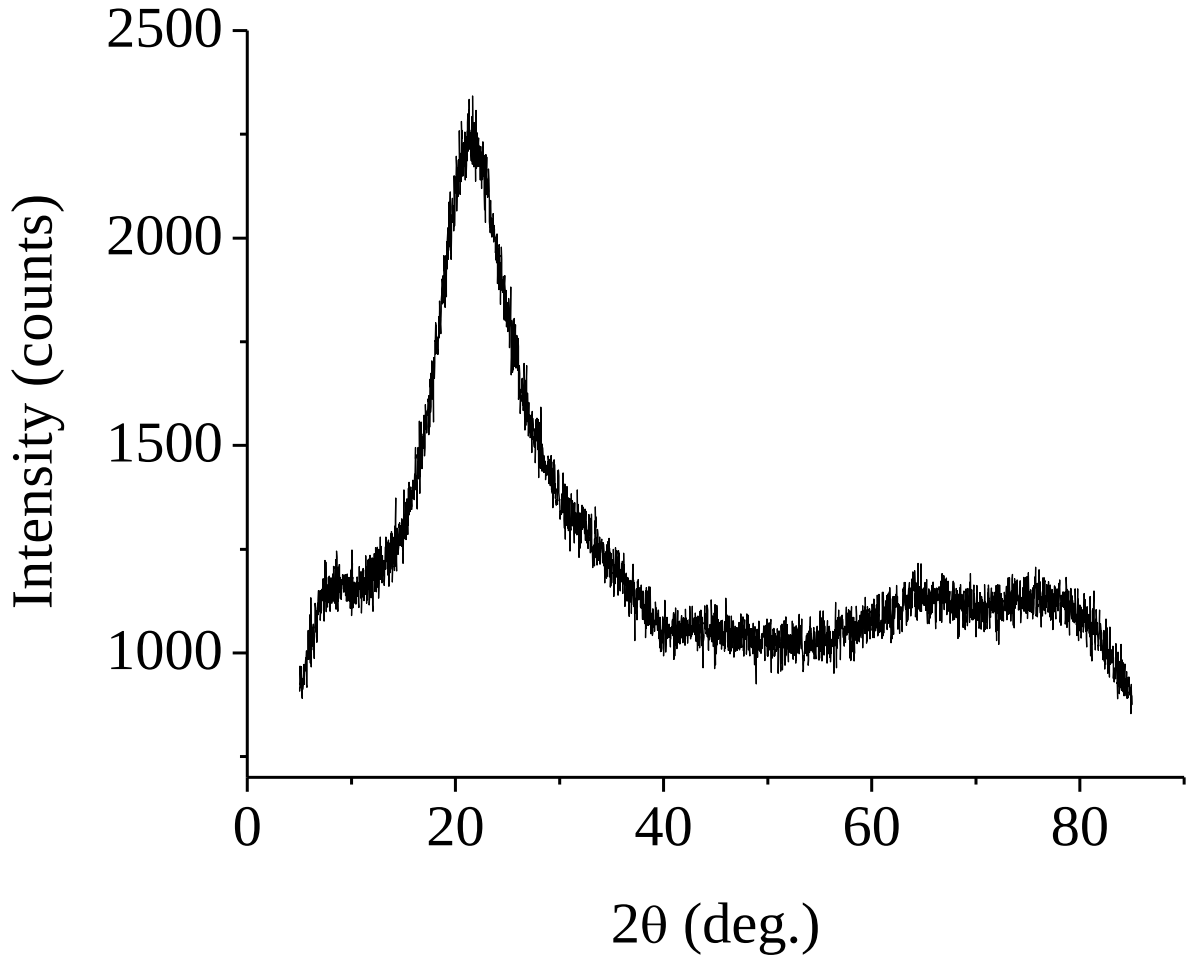

**Figure 5.** XRD of deposited amorphous carbon film on quartz plate.

The nature of the elements present in the film at various places along the axis of the tube is determined using EDX measurement. It shows that film is composed of mostly carbon with traces of O and N elements (also contains hydrogen, on the basis of FTIR-ATR spectra). Oxygen, water and nitrogen molecules are the impurities in the supplied gas mixture, Ar/$CH_4$. Presence of trace of W element in addition to the traces of O and N elements has been observed for the film deposited near the spike. In order to investigate the chemical nature of these elements and their composition, XPS measurements are carried out. It also indicates that the film is composed of mostly carbon with traces of O and N. Chemical nature of the carbon is analyzed using curve fitting procedure (figure 6, left). It shows that film is mostly composed of $sp^2$ carbon bonded to other $sp^2$ carbon with less quantity of $sp^3$ carbon bonded to other $sp^3$ carbon [29]. It also shows that some quantity of carbon is attached to O as well as to N. This fact is also supported by the characteristic peak for O and N in the XPS spectrum.

The film has two phases as shown in figure 6, right. One phase in the film has large area with smooth surface, as also shown in figure 2, right.  The secondary phase in the film has hemispherical top faces, which are occupied in a small area of total film surface. Since the composition of $sp^2$ carbon in the film is higher than the $sp^3$ carbon, it could be that the smooth surface is composed of $sp^2$ carbon, while the secondary phase in the film with a hemispherical top face could be of hydrogenated $sp^3$ carbon material (figure 6, right). Preliminary chemical inertness test shows that deposited film is stable against brushing with water, *iso*-propanal and soap solution.

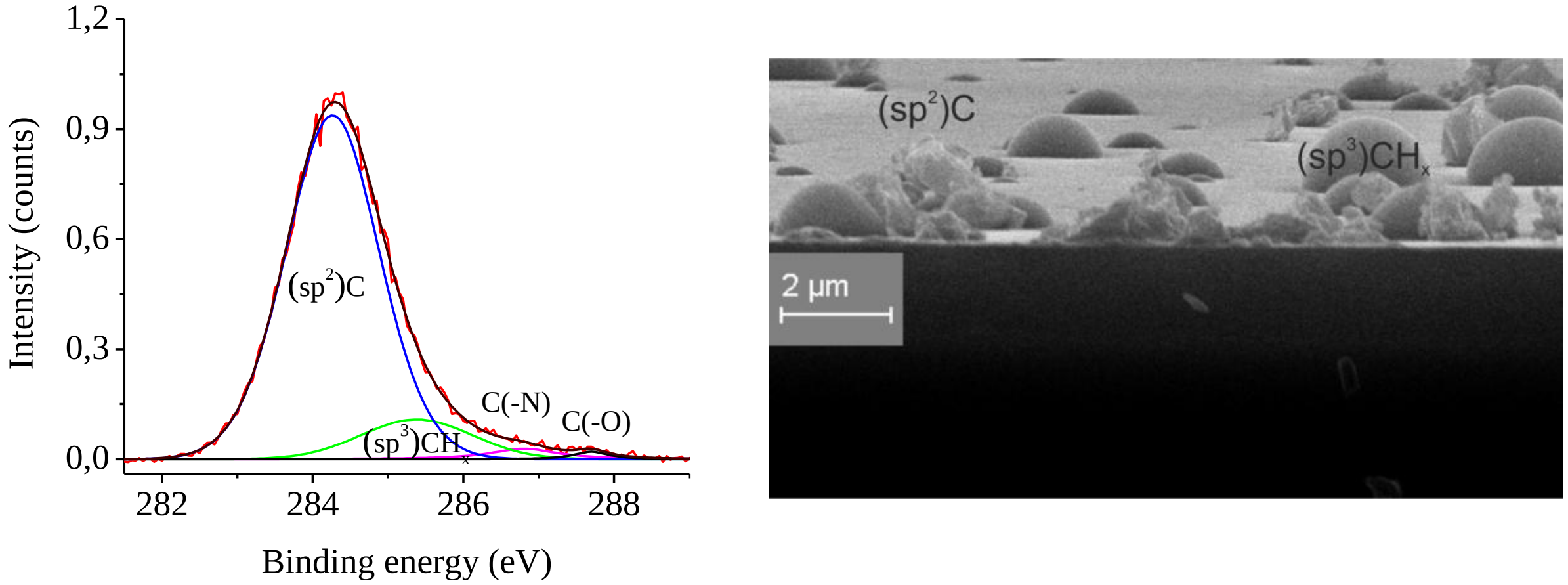

**Figure 6.** XPS of carbon (left) and SEM image (right) of the film at middle region in between electrodes.

*3.2. Characterisation of plasma conditions*

To study the mechanism of film deposition we characterize plasma conditions (gas temperature in plasma, electron density, EVDF, etc.), calculate production rates of atoms and excited molecules, simulate fluxes of excited chemical species to the inner surface of the tube, and simulate the chemical kinetics. OES, microphotography and current-voltage measurements are used for the characterization of plasma conditions. In the following section, gas temperature, reduced electric field and electron density in the plasma are discussed. Because of the high neutral gas temperature in pulsed plasma filament, the tube is also heated. Since the temperature of tube will influence the film growth process and film properties, tube temperature is also determined. It is discussed along with gas temperature in plasma and its fluxes towards surface of tube.

The gas temperature in the plasma filament is about 1000 K. Although it is high (figure 7a), duration of the filamentary discharge pulses is short of about 160 ns. Therefore, the actual (stationary) temperature of tube will be much lower than the gas temperature in the plasma filament. The actual temperature of the tube during the discharge is measured using a thermocouple. It shows that tube temperature is less (about 330 K) close to the spike and it reaches about 400 K close to the grounded electrode (figure 7b). Gas flow could be one of the reasons for this trend in tube temperature, since the gas mixture at room temperature is entering the spike region and relatively hot (or warm) gas mixture is entering the region close to the grounded electrode. With these data, equation for thermal conductivity is numerically solved to simulate the gas temperature in afterglow phase with temporal and spatial resolution.

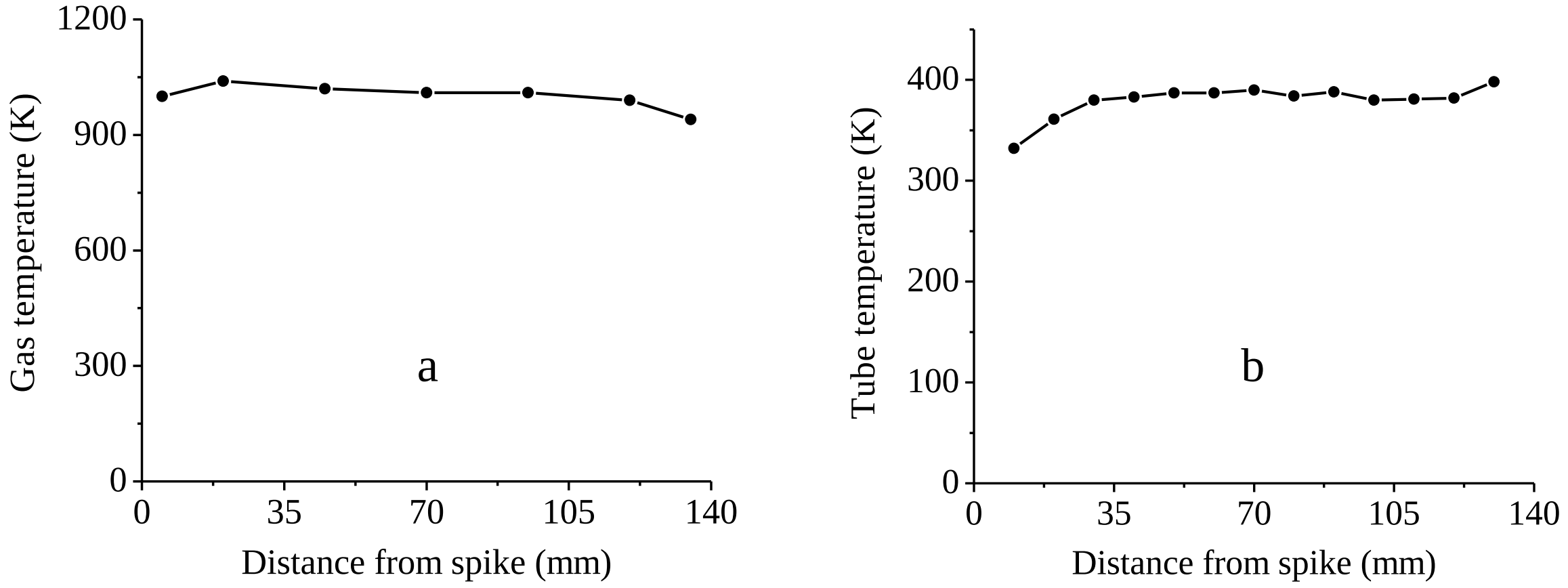

**Figure 7.** Gas temperature in the filament (a) and tube temperature (b) along the axis of the tube during the pulsed positive filamentary discharge (Ar, 99.85%; $N_2$, 0.11%; $CH_4$, 0.04%; total gas flow rate, 2.4 slm).

The results of this simulation presented in figure 8 show that the steady state conditions in the tube will be reached after 500 pulses. It is to be noted that, as mentioned above, about 2000 filaments in series have the same profile and position.

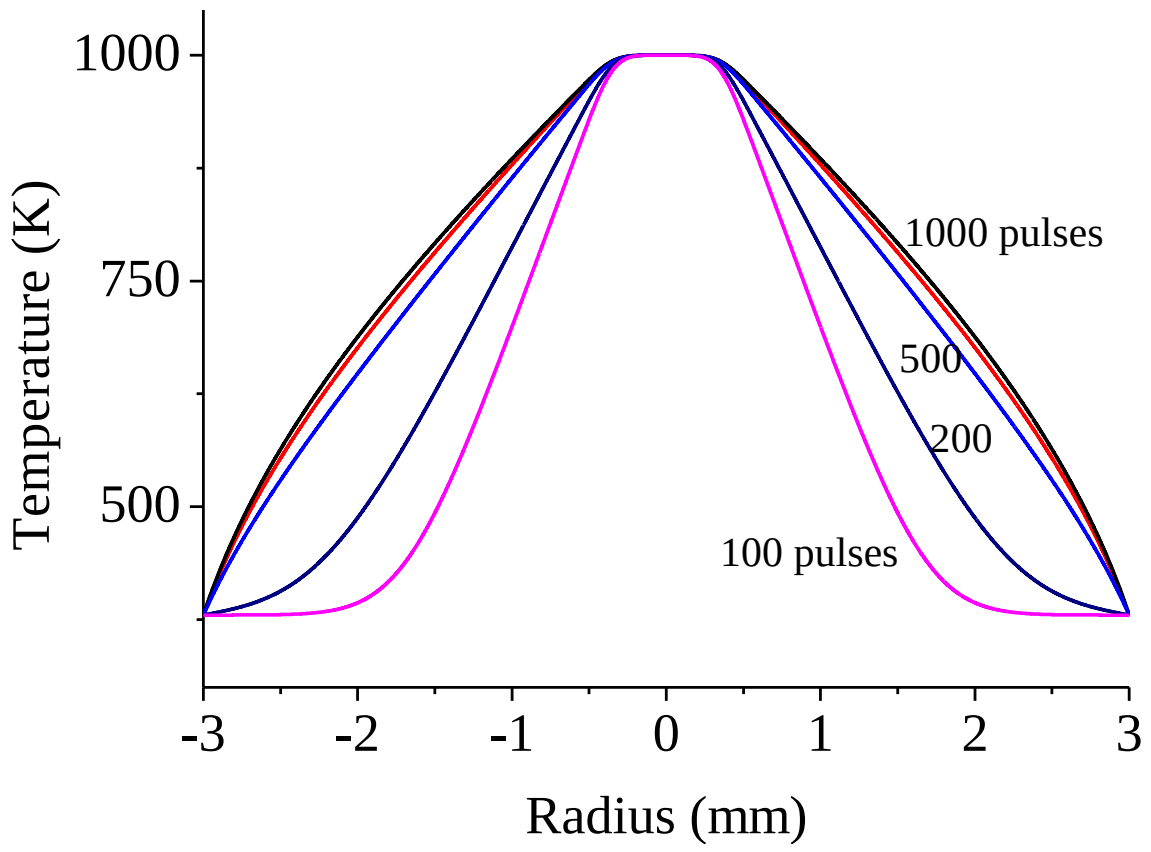

**Figure 8.** Spatial distribution (in the radial direction) of the gas temperature from the filament towards the surface of tube.

The reduced electric field determined (section 2.2) during the positive filamentary discharge along the axis of the tube is shown in figure 9. It reveals that the reduced electric field is higher near the spike of the powered electrode than near the grounded electrode. Standard deviation in the reduced electric field is estimated by considering the fluctuation in the intensity of emissions of $N_2$(C–B, 0–0) and $N_2^+$(B–X, 0–0).

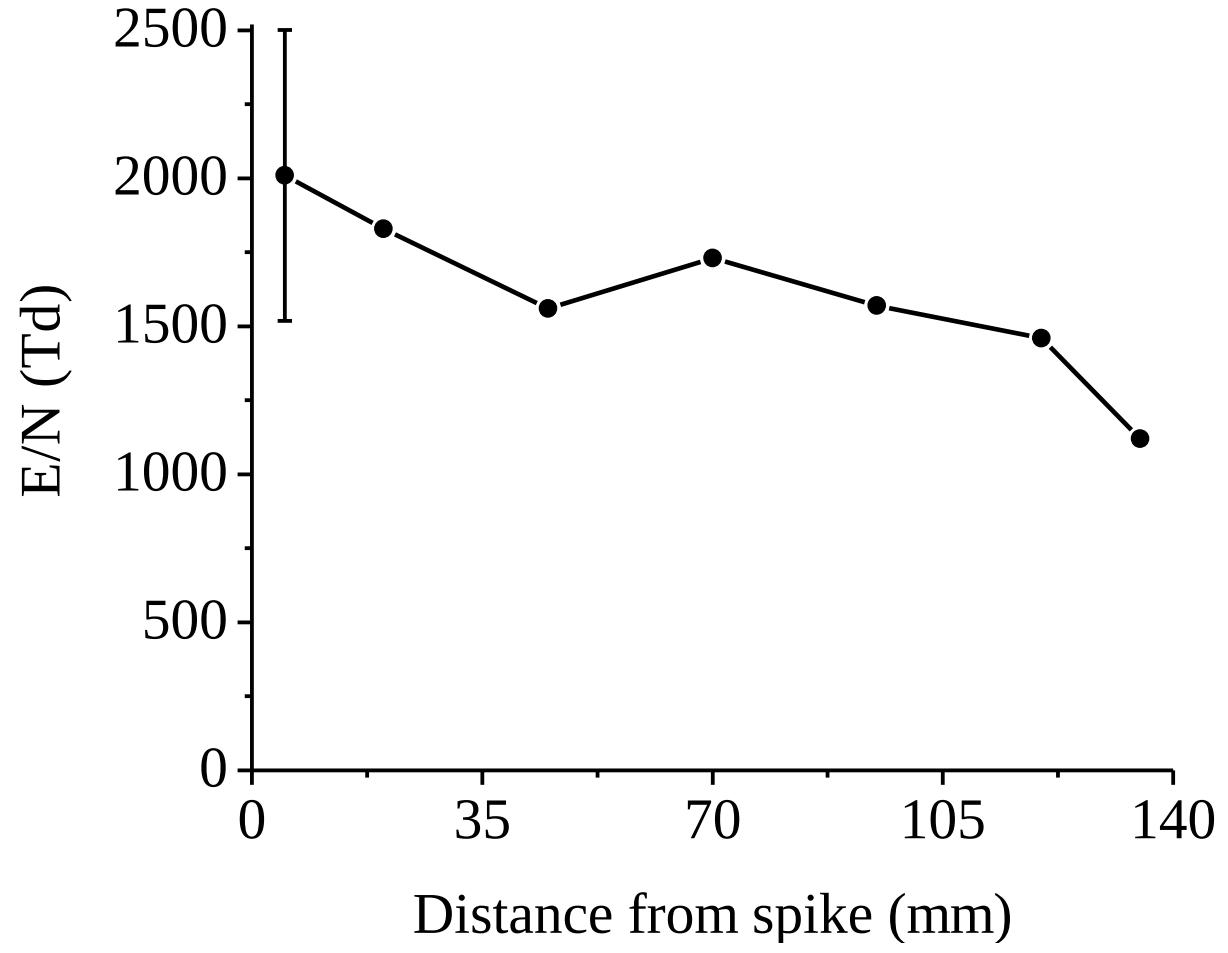

**Figure 9.** Variation of the reduced electric field along the axis of the tube during the pulsed positive filamentary discharge (Ar, 99.85%; $N_2$, 0.11%; $CH_4$, 0.04%; total gas flow rate, 2.4 slm).

The electron density is almost constant ($1.7\times10^{12}$ to $2.8\times10^{12}$) in most part of the region in between the electrodes (figure 10). The filament propagates far outside of the grounded region during the discharge with very less intensity of emission compared to the region in between the electrodes. Investigation of this process is under progress.

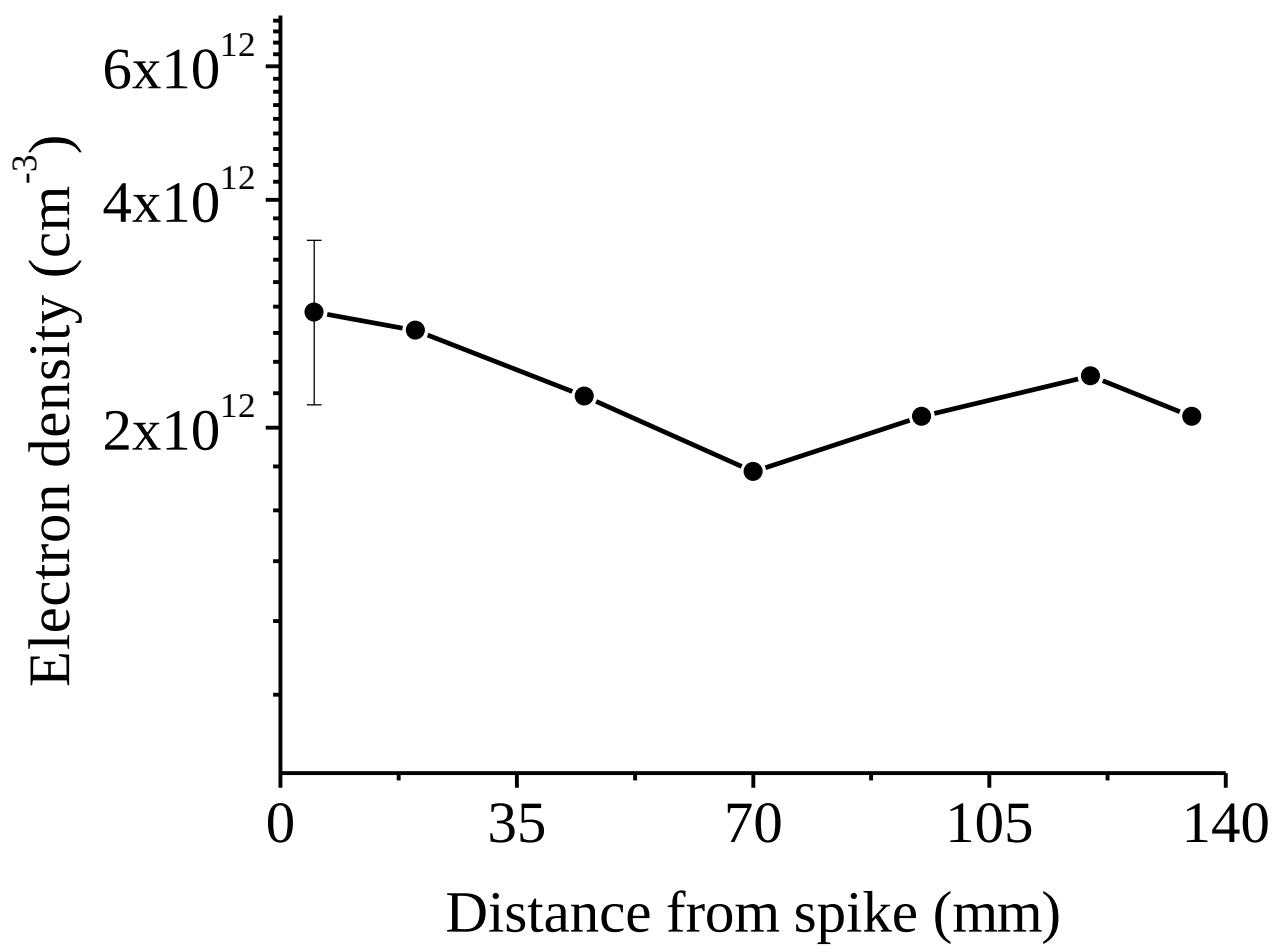

**Figure 10.** Electron density profile along the axis of the tube during the pulsed positive filamentary discharge (Ar, 99.85%; $N_2$, 0.11%; $CH_4$, 0.4%; total gas flow rate, 2.4 slm).

*3.3. Chemical kinetics*

The rate constants for electron impact methane dissociation, argon ionization and argon metastables formation are determined (figure 11) using the equation (1), from the known values of cross-sections for the corresponding process [30-33] and determined EVDFs.

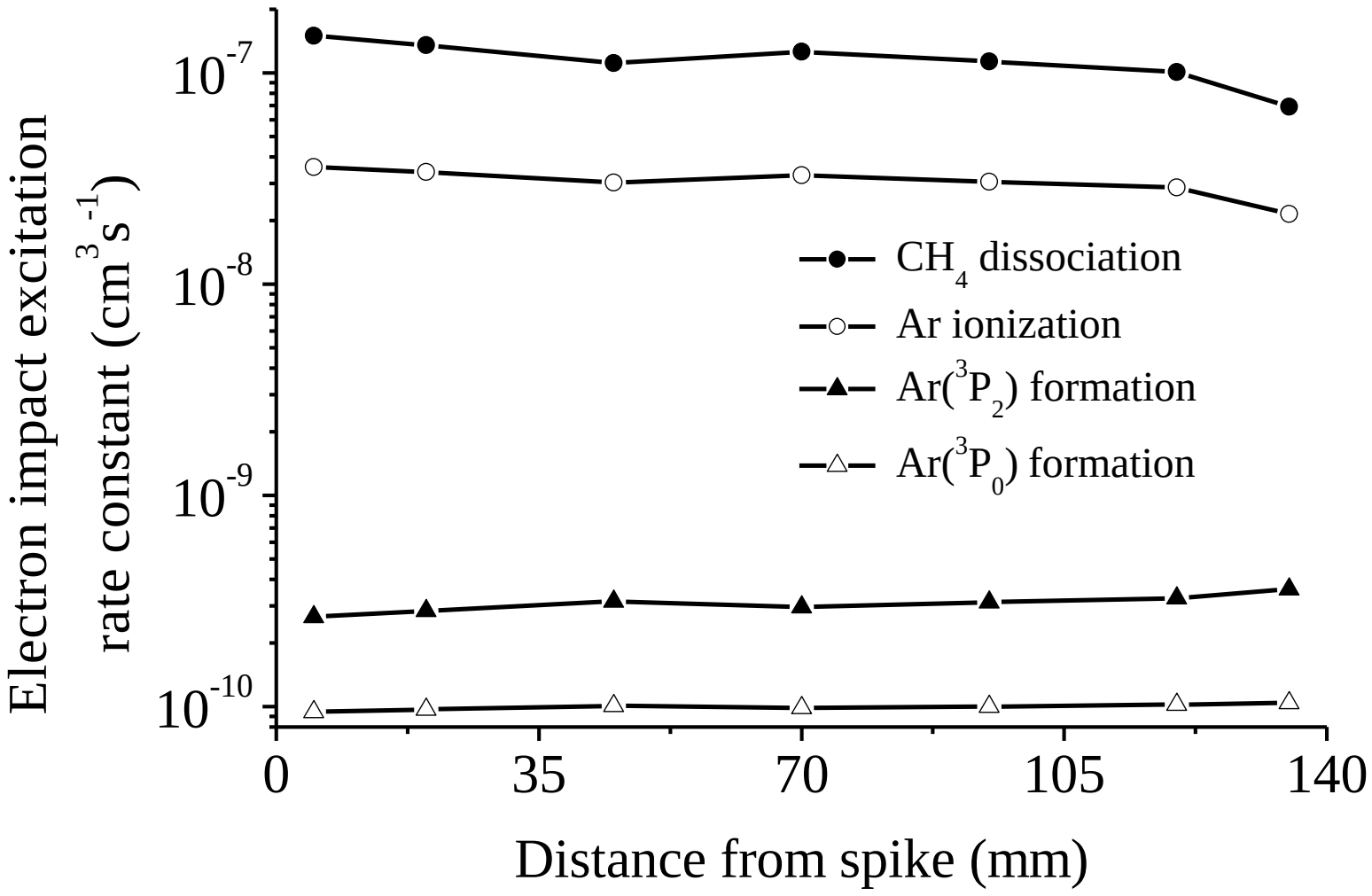

**Figure 11.** Variation in the rate constants for electron impact dissociation of methane, argon ionization and argon metastables formation along the axis of the tube during the pulsed positive filamentary discharge (Ar, 99.87%; $CH_4$, 0.13%; total gas flow rate, 2.4 slm).

Electron impact dissociation rate constant of methane molecules is higher than the electron impact ionization rate constant of argon atoms (figure 11). Despite this fact, because of the higher concentration of argon than methane in the supplied gas mixture, electron impact argon ionization rate is much higher than the electron impact methane dissociation rate. Since the charge exchange reactions of argon ions with hydrocarbon molecules (methane) have big cross sections [34], mainly these reactions are involved in the ionization and dissociative ionization of hydrocarbon species. Hence, argon ion plays dominant role in the methane ionization, dissociation and consequent film deposition processes.

Since the exact density of various chemically active species and their diffusion coefficients are not known, methane diffusion in argon is simulated in order to have some hints about the diffusion of various chemically active species. Simulated results for methane diffusion in the figure 12 show that it takes about 10 to 20 ms for a chemically active species generated in the filament, which is approximately placed at the middle of the tube, to reach the tube surface. During this transport process, it may undergo several chemical reactions. Gas velocity is 140 cm·s$^{-1}$ for the gas flow rate of 2.4 slm. Hence, in the duration of 10 to 20 ms, neutral radicals will be moved to a distance of 1.5 to 3 cm along the axis of the tube.

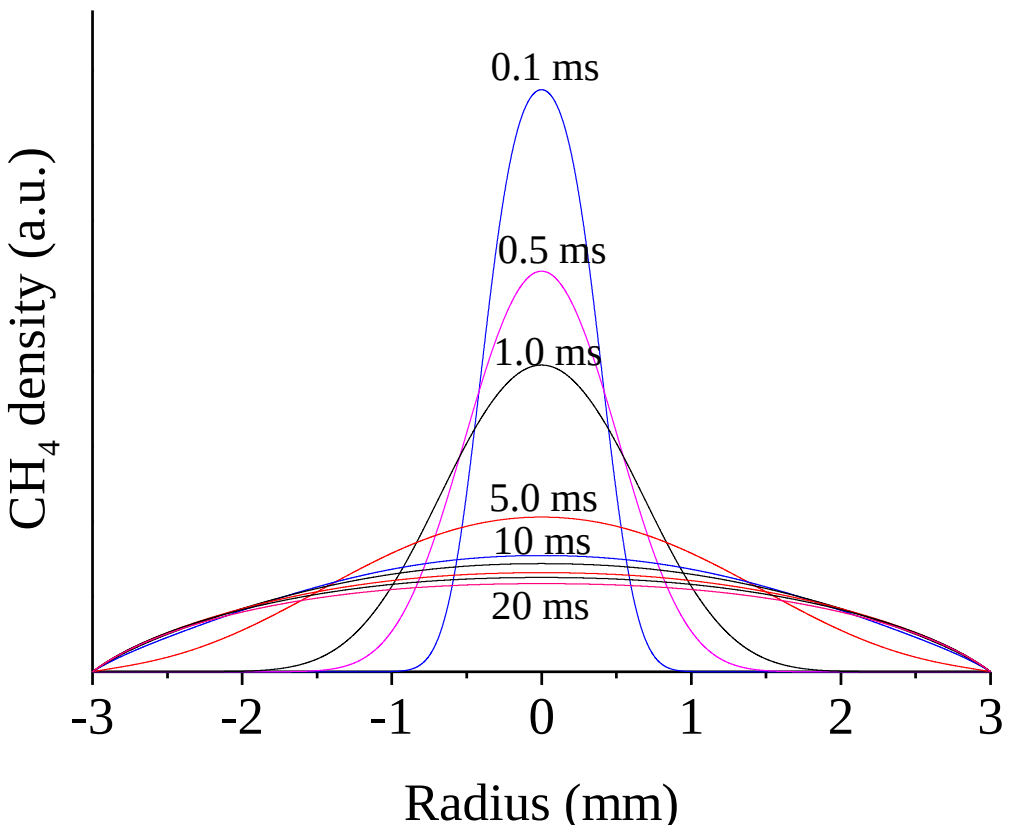

**Figure 12.** Spatial distribution of chemically active species generated in the filament towards the surface of tube. Since several chemically active species are generated in the filament, $CH_4$ diffusion in Ar is simulated.

On the basis of densities of argon, methane and electron, as well as rate constants of various processes determined from the plasma parameters [35-41], most probable reactions among others are calculated using the equation (5),

$$\tau = \mathrm{P}^{-1} = ([\mathrm{M}] \cdot k)^{-1} \qquad (5)$$

where, [M] is the density of argon or methane or electron ($cm^{-3}$), P is the probability and $k$ is the rate constant ($cm^3\ s^{-1}$). The possible important chemical reactions with the inverse of their probability (τ, life time of active species concerning the corresponding reaction) are shown in the following scheme (figure 13). The high probable reactions among others are shown in thick arrows in the scheme. Because of high electric field in active plasma volume, argon ion formation is more probable than argon metastable formation. At atmospheric pressure conditions, $Ar^+$ ion produces molecular ion, $Ar_2^+$ through a three particles reaction. Life time of this molecular ion in argon-methane mixture is very short because of effective charge transfer reaction with methane [34]. This reaction produces $CH_4^+$ ion. This ion undergoes further reaction with $CH_4$ to produce $CH_5^+$ ion and $CH_3$ radical. These chemically active species on further reaction with electrons present in the afterglow phase leads to the formation of $CH_3$ radical. Though this methyl radical is mostly produced through the above stated mechanism, it is also produced by direct reaction of argon metastables with methane as well as by electron impact dissociation of methane. Electron impact dissociation of methyl radical ultimately leads to the formation of carbon atom through $CH_2$ and CH species. The presence of C and CH are confirmed through their characteristic emission (figure 14) [42-44] measured using echelle spectrometer. In addition to the production of molecular argon ion ($Ar_2^+$) through termolecular reaction, argon ion ($Ar^+$) also produces $CH_3^+$ ion through a high probable reaction with methane. The further reaction of $CH_3^+$ ion with methane and electron leads to the formation of diatomic molecular carbon $C_2$, which has been observed through emission spectrum. The $C_2$ molecule on further reaction with electron leads to the formation of atomic carbon and its presence has also been observed through its emission. Since the electron density in afterglow phase is not known, probability for the electron impact reactions in afterglow phase could not be calculated.

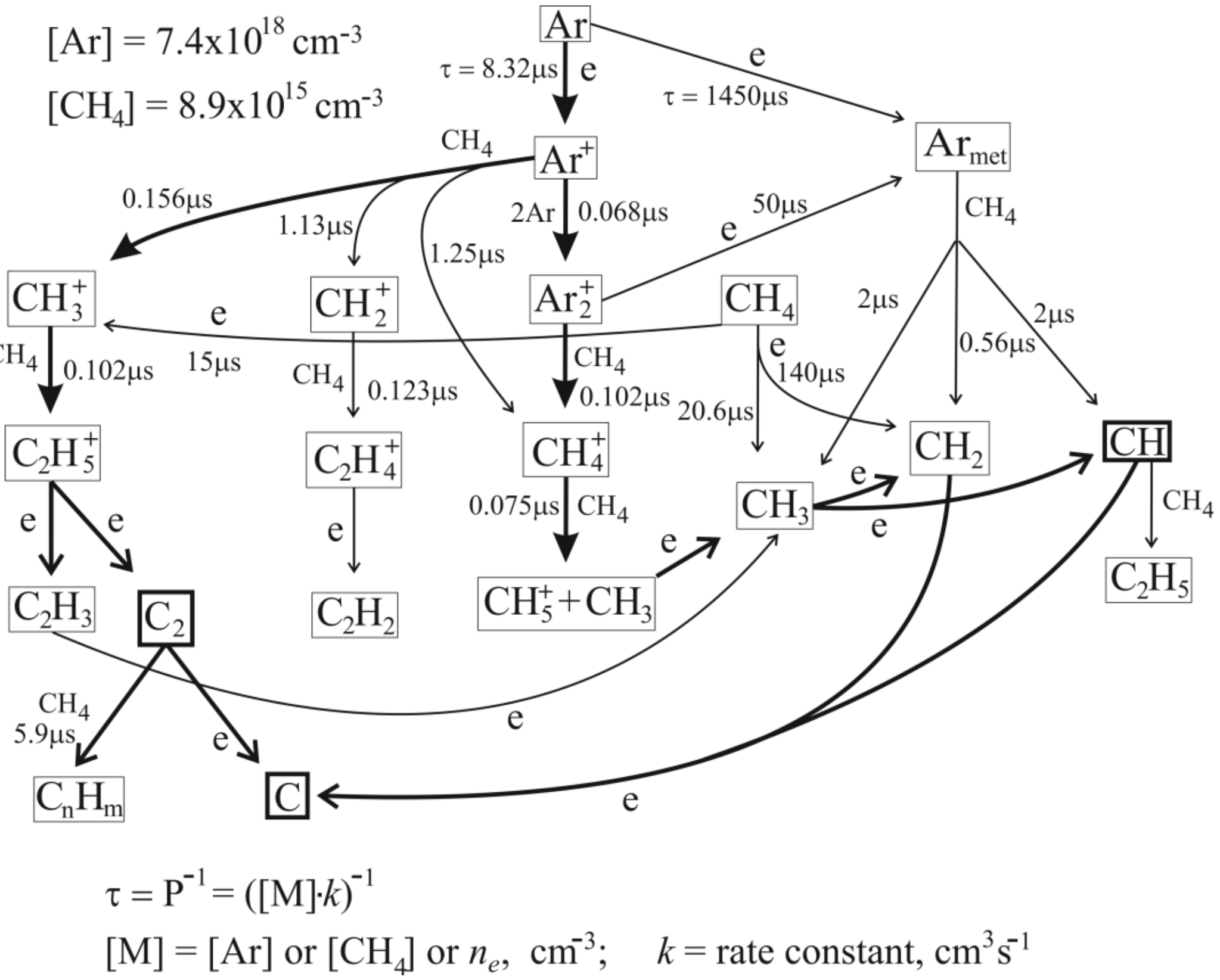

**Figure 13.** Chemical reactions involved in the film growth process. Thick arrow indicates high probable reaction. Emissions of active species in thick frames are observed in the measured emission spectrum.

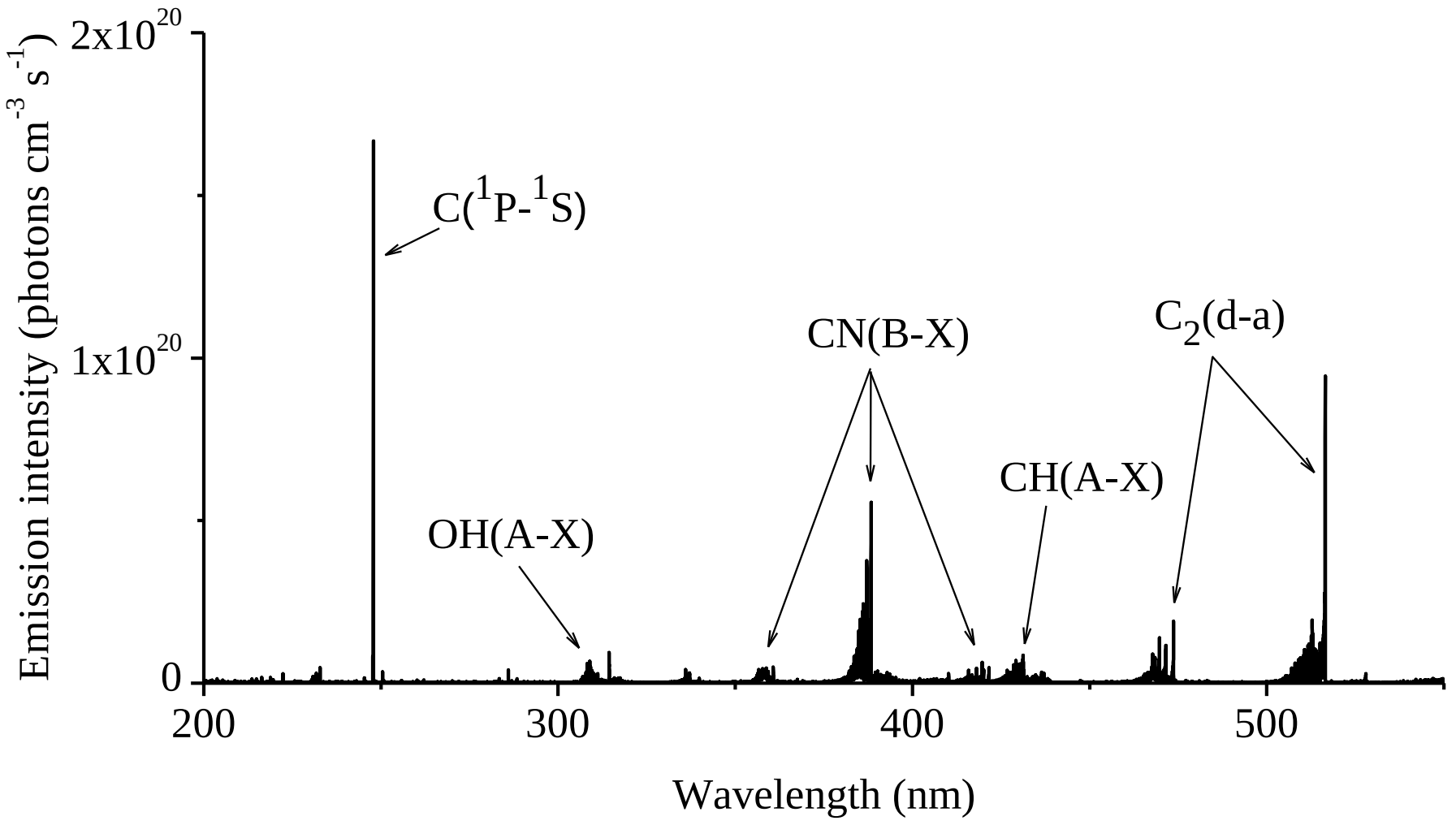

**Figure 14.** Emission spectrum measured at 70 mm from the spike using echelle spectrometer during the pulsed positive filamentary discharge ignited at total gas flow rate of 2.4 slm (Ar, 99.87%; $CH_4$, 0.13%).

To determine life time of positive ions in afterglow phase of filamentary discharge, measured current is compared with displacement current ($I_{disp.}$), which is calculated as $I_{disp.} = dV/dt$ and normalized to the measured current immediately after the positive filamentary discharge (figure 15). The decay time of charged particles in the tube in afterglow phases of filamentary discharge is about 30 μs. Since the measured current is higher than the calculated displacement current in first positive and first negative phases of voltage following the ignition of filamentary discharge, it is concluded that some additional active discharge is ignited in the tube after the positive filamentary discharge. Microphotographical studies of this phenomenon show that it is a glow discharge, because the “negative glow” covers the surface of electrode when it is cathode. Electric field and photoemission in this discharge phase are low, and hence it does not influence considerably on the plasma parameters determined using OES for filamentary discharge.

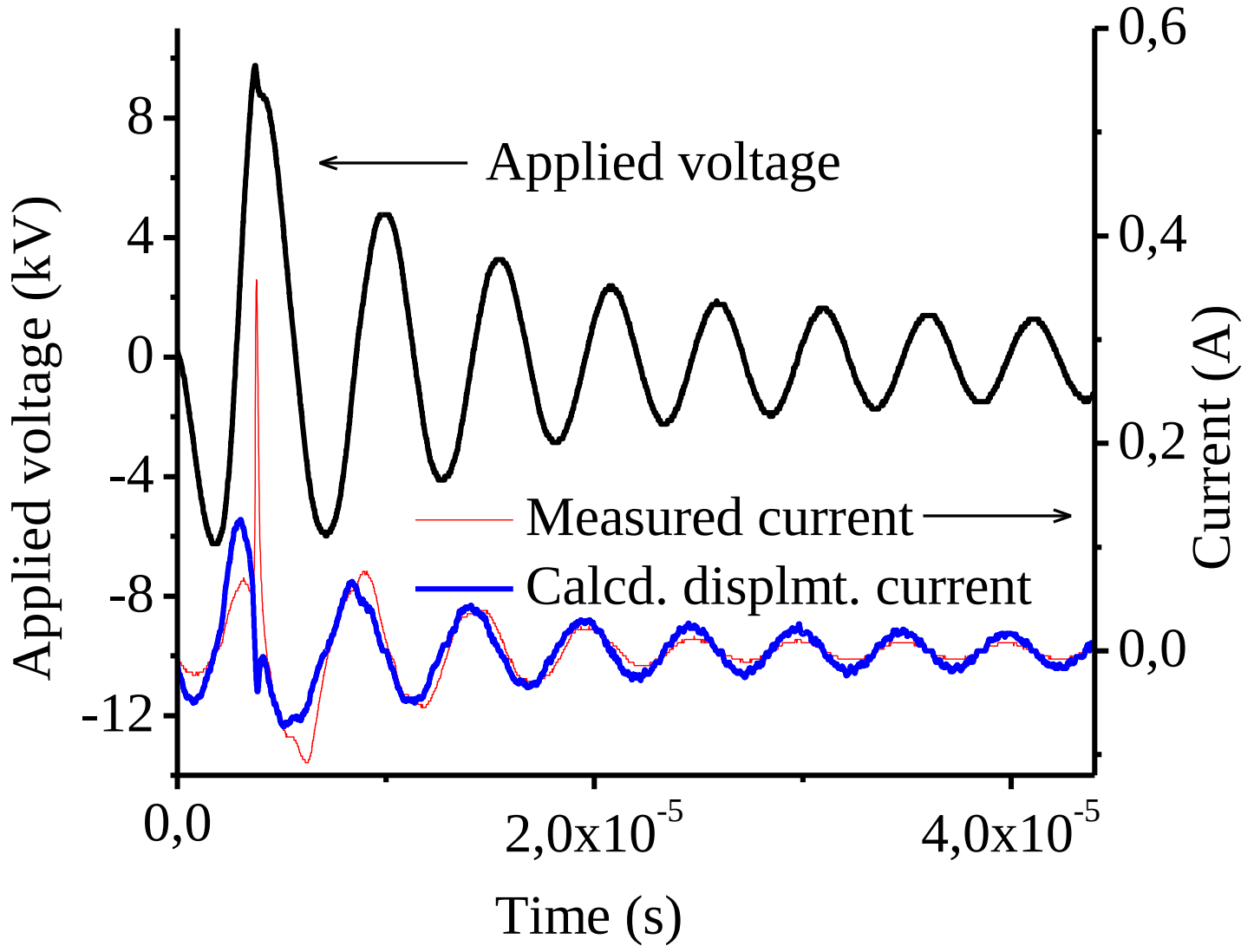

**Figure 15.** Voltage and current profiles during the pulsed positive filamentary discharge (Ar, 99.87%; $CH_4$, 0.13%; total gas flow rate, 2.4 slm). The pulse frequency is 22 kHz. The voltage frequency of each pulse sequence is around 200 kHz. Displacement current is calculated as $I_{disp.} = dV/dt$ and normalized to the current measured immediately after the positive filamentary discharge.

### *3.4. Mechanism of film deposition*

On the basis of i) plasma parameters determined applying optical emission spectroscopy and simulation, ii) rate constants calculated using plasma parameters, iii) voltage-current measurements and iv) characterization of the film deposited inside the tube between spike and grounded area, it is concluded that the hydrocarbon ions produced in the charge exchange reactions of argon ions with hydrocarbon molecules play a dominant role for film deposition. At steady state conditions, methane, which is a part of gas mixture flows through the tube, is fully ionized and dissociated in the tube at the distance of approximately 20 mm from the spike. However, the film deposition rate has sharp maxima under the grounded area, which is at 140 mm away from the spike. The film deposited in the area between the spike and the grounded electrode has different thickness at different places, but with similar properties and components.  This film is dense and amorphous; has smooth surface and low content of hydrogen. We suppose that this film is deposited during ion bombardment on the wall after ambipolar diffusion and drift. In the frame of this assumption, we can explain different observed effects such as i) increase of deposition rate in the region where filament passes on or near the wall because of small distance to the wall, ii) higher deposition rate in the region under grounded electrode because of higher drift velocity in comparison to the velocity of ambipolar diffusion, and iii) low deposition rate near the grounded area because of high axial component of electric field in this region. The hydrocarbon film deposited in the region beyond the grounded area is differed strongly from the film deposited before and under the grounded area. This film is soft, rough and low dense. This film may be deposited by flux of neutral hydrocarbon species which are produced by collisions of hydrocarbon ions with the surface of the tube under the grounded area. The kinetic energy of these ions is high enough for their partial dissociation. The neutral hydrocarbon species formed in this process flow with gas along the tube and deposit on the wall after diffusion.

The mechanism of ions transport from the spike area to the grounded area is under investigation. The steady state conditions of this filamentary discharge depend on the distance between the electrodes as well as the amplitude and the frequency of applied voltage. For longer distance between electrodes, lower voltage frequency requires higher applied voltage for steady state discharge. We suppose that the residual charged species produce the electric field near the propagated head of the filament, and stimulate the propagation to a long distance. As mentioned above, the life time of the ions in the tube is about 30 μs, which is comparable with the pulse duration (45 μs). During this period, the charged species diffuse from the plasma channel. The average density of these charged species during the initiation of next discharge is about $10^9$ $cm^{-3}$ at the steady state conditions. On the other hand, very high density of the charged species also hampers ignition of the filamentary discharge, as discussed in the previous publication [12].

The facts discussed above show that plasma conditions in our experiment are different along the plasma channel. To make plasma conditions more uniform along the axis of the tube, the applied voltage with frequency of several tens of kHz can be modulated with sufficiently low frequency (pulse package modulation frequency) for filling of all tube with fresh gas mixture between plasma treatment phases. These experimental conditions will be used and studied in our forthcoming investigations.

## 4. Summary

The pulsed filamentary plasma source has been constructed to produce a long filament of plasma inside a tube for film deposition on inner surface of tubes. FTIR-ATR, XRD, SEM, LSM and XPS analyses give the conclusion that deposited film is amorphous composed of non-hydrogenated $sp^2$ carbon and hydrogenated $sp^3$ carbon with traces of O and N. Smooth film in the SEM image is composed of $sp^2$ carbon while the secondary phase in the film with the hemi-spherical top face is composed of hydrogenated $sp^3$ carbon. UV-Vis absorption spectra of the film deposited on inner surface of the tube also confirm the presence of doubly bonded $sp^2$ carbon, which are shortly conjugated. Optical emission spectroscopy, voltage current measurement, microphotography and numerical simulations are used for determination of the plasma parameters and the gas temperature. With these data, equation for thermal conductivity is numerically solved to simulate the gas temperature in afterglow phase with temporal and spatial resolution. The rate constants for argon ionization and methane dissociation are almost constant in the entire region in between electrodes although it is slightly higher close to the spike. Based on the analysis of kinetics of deposition processes, it is observed that ions play important role in the deposition of amorphous carbon film.

## Acknowledgement

This work is supported by the ‘Deutsche Forschungsgemeinschaft’ (DFG) within the frame of the research group ‘FOR1123 - Physics of Microplasmas’. We thank Dr. Kirill Yusenko for XRD spectra.